# Basis Path Coverage Criteria for Smart Contract Application Testing

Xinming Wang
*School of Computer Science*
*South China Normal University*
Guangzhou, China
wangxinming@m.scnu.edu.cn

Zhijian Xie
*School of Computer Science*
*South China Normal University*
Guangzhou, China
xiezhijian@m.scnu.edu.cn

Jiahao He
*School of Computer Science*
*South China Normal University*
Guangzhou, China
hejiahao@m.scnu.edu.cn

Gansen Zhao[§]
*School of Computer Science*
*South China Normal University*
Guangzhou, China
gzhao@m.scnu.edu.cn

Ruihua Nie
*School of Computer Science*
*South China Normal University*
Guangzhou, China
nrh@scnu.edu.cn

*Abstract*— The widespread recognition of the smart contracts has established their importance in the landscape of next generation blockchain technology. However, writing a correct smart contract is notoriously difficult. Moreover, once a state-changing transaction is confirmed by the network, the result is immutable. For this reason, it is crucial to perform a thorough testing of a smart contract application before its deployment. This paper's focus is on the test coverage criteria for smart contracts, which are objective rules that measure test quality. We analyze the unique characteristics of the Ethereum smart contract program model as compared to the conventional program model. To capture essential control flow behaviors of smart contracts, we propose the notions of *whole transaction basis path set* and *bounded transaction interaction*. The former is a limited set of linearly independent inter-procedural paths from which the potentially infinite paths of Ethereum transactions can be constructed by linear combination, while the latter is the permutations of transactions within a certain bound. Based on these two notions, we define a family of path-based test coverage criteria. Algorithms are given to the generation of coverage requirements. A case study is conducted to compare the effectiveness of the proposed test coverage criteria with random testing and statement coverage testing.

## I. INTRODUCTION

Blockchains as one form of distributed ledger technology have gained considerable interest and adoption since Bitcoin was introduced [1]. Participants in a blockchain system run a consensus protocol to maintain and secure a shared ledger of data (the blockchain). Blockchains were initially introduced for peer-to-peer payments [1], but more recently, it has been extended to allow programmable transactions in the form of *smart contracts* [2], [3]. Smart contracts are programs that can be collectively executed by a network of mutually distrusting nodes, who implement a consensus protocol (such as proof-of-work [1] or proof-of-stake [5]) that digitally enforce agreements among nodes: neither the nodes nor the creator of the smart contract can feasibly modify its code or subvert its execution. Ever since being proposed and implemented by blockchain systems such as Ethereum [3] and EOS [6], smart contracts have been applied across a range of industries, including finance, insurance, identity management, and supply chain management. In our paper, we will assume Ethereum as the smart contract platform.

Because smart contracts are entrusted by the users to handle and transfer assets of considerable value, they are subject to intensive hacking activities. Such hacking is more dangerous than that on a conventional network system, because once deployed on the blockchain, the contract becomes immutable, essentially creating a high-risk, high-stake paradigm: the deployed code is nearly impossible to patch, and contracts collectively control billions of USD worth of digital assets. For example, there have been a plenty of well-documented attacks on the Ethereum smart contracts [7]: The reentrancy attack managed to steal tokens valued $60M from a contract and ultimately led to the hard-fork that created Ethereum Classic (ETC) [8]. The Second Parity Multisig Wallet hack exploits well-written library code to run it in non-intended context [9]. About $300M-worth cryptocurrency was frozen and (probably) lost forever. In April 2018, BecToken was attacked due to integer overflow on multiplication, causing an extremely large amount of tokens transferred to malicious accounts and the token price dropped to nearly zero [10]. The severe consequence of these attacks highlights the importance of verification before smart contracts deployment.

In the literature, a number of approaches and tools have been proposed to verify smart contracts [11]–[16]. Most of them focus on detecting *vulnerabilities*, which are known code patterns that have been reported previously. Examples include re-entrancy, unchecked send, arithmetic overflow, and dangerous delegatecall [7]. Although vulnerability detection tools have been shown to be effective at detecting known dangerous code patterns, they are far from enough to verify the correctness of smart contracts. This is because not all smart contract defects involve known code patterns. Some, instead, are specific to the application logic of the particular contract under concern. These defects are referred to as *logic error* in

---

[§] Correspondence author.

[*] This work was supported by Science and Technology Planning Project of Guangdong Province, grant Nos. 2016B030305006，2017KZ010101，2017B030308009，and 2018A070717021.

```
1: function checkNoExists(uint id) returns(bool){
2:     for (uint i = 0; i < numbers.length; i++){
3:         if (numbers[i] == id) {
4:             return true;
5:         } else{ // Shall put outside the loop
6:             return false;
7:         }
8:     }
9: }
```

Figure 1: A logic error in the smart contract reported at [14].

the empirical investigation by Kalra and colleagues [14]. For example, consider the real world example illustrated in Figure 1. By mistake, the smart contract programmers put "`return false;`" at line 6 instead of line 9 after the loop. Logic errors of this kind cannot be detected by existing vulnerability detection tools, because no consistent code patterns can be summarized from them. Another reason why existing vulnerability detection tools might be insufficient is that there can be still plenty of unknown vulnerabilities that do not match any patterns reported by practitioners and researchers.

For the above reasons, we believe that beside vulnerability detection, testing is still indispensable to guarantee the correctness of smart contracts. Note that the need of more research on systematic smart contract testing has already been advocated by several position papers [17], [18]. However, to the best of our knowledge, this direction has been largely overlooked in the research community. By contrast, the importance of testing is well recognized by smart contract developers. For instance, as one of the most widely used testing tools, Truffle suite [19] allows developers to encode test inputs and assertions into test scripts using JavaScript language. These test scripts are then executed on a blockchain simulation environment (Ganache). Such test frameworks, however, cannot guide the testing process in the sense that they do not tell the developers what kind of test scripts they should write, and whether existing test cases are sufficient. To address this issue, *test coverage criteria* [20] for smart contracts need to be defined.

Test coverage criteria [21]–[23] are sets of rules that help determine whether a test suite has adequately tested a program. Many test coverage criteria have been proposed and empirically compared for conventional programs, including control-flow coverage criteria, such as branch-coverage [24], and data-flow coverage criteria, such as all-def-use-pairs [25]. Smart contracts, however, have several unique characteristics that can make existing test coverage criteria inappropriate. Firstly, the execution of smart contracts are organized around units know as *transactions*. The interaction between transactions across multiple contracts are crucial in the effects on the contract state. Therefore, test coverage criteria shall be designed according to the categorization of such interactions in practice, which is a factor not considered in conventional test coverage criteria. Secondly, within each transaction, the control flow can be transferred in an unexpected way. The best example is the famous DAO attack [8], which is caused by the unexpected reentrancy of the contract function. Conventional test coverage criteria are insufficient to address whole transaction control flow, because most of them are intra-procedural. While a few works do propose inter-procedural coverage criteria (e.g. [26], [27]), they only focus on the interfaces between functions.

The above characteristics suggest the need to develop test coverage criteria based on how the execution of individual smart contract transaction traverses across contracts, and how transactions interact with each other in a sequence. The key challenge, however, is that both aspects involve potentially infinite number of coverage requirements. The number of possible execution paths of a transaction can be infinite when the contract code contains loops, and transactions sequence can also be extended infinitely. Therefore, directly defining test coverage criteria on them will make coverage testing of smart contracts intractable.

In order to address this challenge, we propose the notion of *whole transaction basis path set*, which is a limited set of linearly independent inter-procedural paths from which all possible execution paths of transactions can be constructed by linear combination. The main novelty of this notion is that it extends the conventional definition of basis path by McCabe [28], which is for intra-procedural testing, to the inter-procedural case. Besides, we also propose the notion of *bounded transaction interactions*, which are all possible permutations of transactions within a certain bound, linked by input/output. Based on these two notions, we define a family of path-based test coverage criteria for smart contract. Algorithms are given to the generation of whole transaction basis paths. A case study illustrates the effectiveness of the test coverage criteria using real world smart contracts with seeded logic defects.

This paper makes the following main contributions:
1) The proposal of practical path-based test coverage criteria for systematic testing of smart-contracts.
2) The algorithms that address the enumeration of coverage requirements.
3) Case study on a real world smart contract application to investigate the effectiveness of the proposed test coverage criteria.

The rest of this paper is organized as follows. Section II introduces the structure of smart contracts and define concepts. Section III define test coverage criteria based on these concepts. Section IV presents algorithms. Section V reports our case study result. Section VI discuss related work. Finally, Section VII concludes.

## II. STRUCTURE AND DEFINITIONS FOR SMART CONTRACT

In this section, we first briefly present the program model of smart contract application on Ethereum. We then introduce the notions of *whole transaction basis path set* and *bounded transaction interaction*. A whole transaction basis path captures the control flow behavior inside one single transaction, while bounded transaction interactions captures the interaction between transactions.

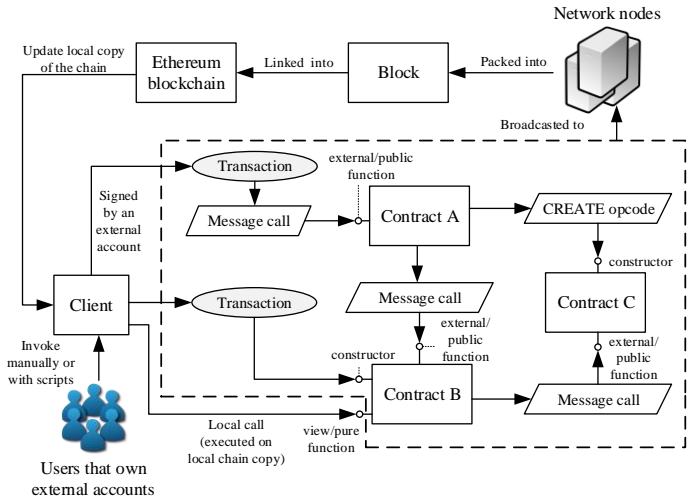

Figure 2: Smart contract program model

*A. Ethereum Smart Contract Program Model*

Figure 2 shows an overview of the model. At the minimal, a smart contract application on Ethereum is composed of one or more *contracts* deployed on the blockchain, and one or more *external accounts* that are owned by the users and hold Ethers as the cryptocurrency. The application relies on *clients* that synchronize with the network to obtain the latest blockchain contents and *nodes* that run the consensus protocol to pack transactions into the blockchain.

Contracts are mostly developed in Solidity [32], which is a JavaScript-like, but typed, programming language. At the source level, contracts written in Solidity appear similar to classes in object-oriented languages. Figure 3 (a) give an example. Contracts can contain declarations of state variables, definitions of functions, modifiers, constructors, structure, and etc. Functions support encapsulation (visibility attributes) and can be statically overloaded. Contracts support interface and multi-inheritance in the C3 linearization style [15]. To enable deployment on the Ethereum platform, the contract functions are compiled into the Ethereum virtual machine (EVM) bytecode and a piece of code called function selector is added, which serves as an entry point into the contract code. Whenever a function is called, the contract code starts executing at the function selector. The selector decodes the message and jumps into an appropriate contract function. Ethereum smart contracts support the custom handling of messages that do not specify a concrete function to call through the fallback function.

After compilation, there are two ways to deploy (also referred to as create) a contract onto the Ethereum network. The first way is by a client sending a *contract creation transaction* with the deployment instruction, bytecode, and paying the creation fee in ether. The transaction is signed by the private key of an external account owned by the user, broadcasted to the network, and packed into a block. The contract instance is deployed as the result. The second way is by another contract executing a special EVM instruction CREATE (keyword **new** in Solidity). In both ways, the constructor will be executed to initialize the created contract.

After deployment, contracts can be invoked with its external/public functions in three different ways. The first way is by the client sending a *message call transaction*, which makes a message call that contains target function selector and parameter data. Similar to the contract creation transaction, the message call transaction must be signed, broadcasted, and packed to take effect. It is a write-operation that will possibly affect other accounts, update the contract state and consequently, the state of the blockchain, and cost Ether. The second way is by another contract directly making a message call to invoke the function. This action is always transitively triggered by a message call transaction. In fact, every message call transaction consists of a top-level message call which in turn can create further message calls. Finally, the third way to interact with contract is by the client locally calling view/pure functions, which never modify the contract state. This kind of calls do not broadcast or publish anything on the blockchain. It is a read-only operation and will not cost any Ether.

One important thing is that the message call transaction is always asynchronous, that is, after the client sends it to a node, the node is not obligated to execute it immediately. When the transaction does get executed, the result will consist of the outcome (success or revert) and execution logs. The client monitors the latest blocks to retrieve this result.

From the perspective of testing, the core model of a smart contract program can be formalized as follows:

A contract application *dapp* is a tuple <*A, C*>, where *A* is a set of external accounts and *C* is a set of contract. At the interface level, each contract $c \in C$ can be denoted as a 5-tuple $< F_P, F_V, f_c, f_d, e >$, where $F_P$ is a set of public/external state-changing functions, which can be called by external accounts $a \in A$; $F_V$ is a set of public/external pure/view functions, which can be directly called on *c* without any external account; $f_c$ is the constructor function of *c*; $f_d$ is the fallback function of *c*; and *e* is the ether value held in *c*. A function *f* is defined as a 3-tuple $<m, (a_0, a_1, \ldots, a_N), (r_0, r_1, \ldots, r_M)>$, where *m* is the function name, $a_i$ is the *i*-th parameter, and $r_i$ is the *i*-th return result. Each parameter or return result can be of primitive type, structured type, function type, or a special type called **address**, which can represent any external account or contract.

For a *dapp* = <*A, C*>, a transaction *T* is defined as a 8-tuple <*a, c, f, e, o, I, R, L*>, where $a \in A$ and $c \in C$. Let $c=< F_P, F_V, f_c, f_d, E>$, either $f = f_c$, $f = f_d$, or $f \in F_P$. If $f = f_c$, the transaction *T* is a contract creation transaction, otherwise *T* is a message call transaction. Symbol *e* is the Ether value sent by the client with the transaction, which is used to pay for the execution fee or send to *c*; *o* is the execution outcome (success/fail); $I = (a_0, a_1, \ldots, a_N)$ represents values of *f*'s *N* parameters ; $R = (r_0, r_1, \ldots, r_M)$ represents values of *f*'s *M* return results; and *L* is a set of string recorded during the execution as logs.

```
contract FishToken{
    ...
    address public currentShark;
    mapping(address => uint256) public balances;
    address[] public participants;
    ...

    function determineNewShark() internal {
1:      address shark = participants[0];
2:      for (uint i=1;
3:           i < participants.length;
4:           i++) {
5:        if (balances[shark]<balances[participants[i]]){
6:          shark = participants[i];
7:        }
8:      }
9:      if(currentShark != shark) {
10:       currentShark = shark;
11:     }
    }

    function addToParticipants(address addr) internal ...{
12:     if(participants[addr]!=null) {
13:       return false;
14:     }
15:     participants.push(addr);
    }

    function transfer(address to, uint256 value) ... {
        ....
16:     addToParticipants(to);
17:     balances[msg.sender] -=value;
18:     balances[to] += value;
19:     determineNewShark();
    }

    function issueTokens(address be, uint256 amount)... {
        ....
20:     addToParticipants(be);
21:     balances[beneficiary] += amount;
22:     determineNewShark();
    }
}
```

(a) code excerpt

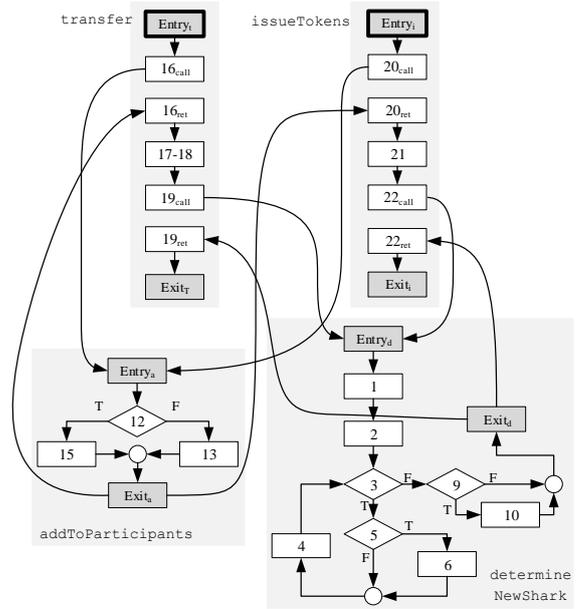

(b) Transaction control flow graph of Pool-Shark

Whole transaction basis path set for `transfer`:
$(_t\text{-}16_{call}\text{-}(_a\text{-}12\text{-}15\text{-})_a\text{-}16_{ret}\text{-}17\text{-}18\text{-}19_{call}\text{-}(_d\text{-}1\text{-}2\text{-}3\text{-}9\text{-})_d\text{-}19_{ret}\text{-})_t$
$(_t\text{-}16_{call}\text{-}(_a\text{-}12\text{-}13\text{-})_a\text{-}16_{ret}\text{-}17\text{-}18\text{-}19_{call}\text{-}(_d\text{-}1\text{-}2\text{-}3\text{-}9\text{-})_d\text{-}19_{ret}\text{-})_t$
$(_t\text{-}16_{call}\text{-}(_a\text{-}12\text{-}13\text{-})_a\text{-}16_{ret}\text{-}17\text{-}18\text{-}19_{call}\text{-}(_d\text{-}1\text{-}2\text{-}3\text{-}5\text{-}4\text{-}3\text{-}9\text{-})_d\text{-}19_{ret}\text{-})_t$
$(_t\text{-}16_{call}\text{-}(_a\text{-}12\text{-}13\text{-})_a\text{-}16_{ret}\text{-}17\text{-}18\text{-}19_{call}\text{-}(_d\text{-}1\text{-}2\text{-}3\text{-}5\text{-}6\text{-}4\text{-}3\text{-}9\text{-})_d\text{-}19_{ret}\text{-})_t$
$(_t\text{-}16_{call}\text{-}(_a\text{-}12\text{-}13\text{-})_a\text{-}16_{ret}\text{-}17\text{-}18\text{-}19_{call}\text{-}(_d\text{-}1\text{-}2\text{-}3\text{-}5\text{-}6\text{-}4\text{-}3\text{-}9\text{-}10\text{-})_d\text{-}19_{ret}\text{-})_t$
$|E'| = 35 \quad |V'| = 32 \quad \text{Cyclomatic Number} = 35\text{-}32\text{+}2 = 5$

Whole transaction basis path set for `determineNewShark`:
$(_t\text{-}20_{call}\text{-}(_a\text{-}12\text{-}15\text{-})_a\text{-}20_{ret}\text{-}21\text{-}22_{call}\text{-}(_d\text{-}1\text{-}2\text{-}3\text{-}9\text{-})_d\text{-}22_{ret}\text{-})_t$
$(_t\text{-}20_{call}\text{-}(_a\text{-}12\text{-}13\text{-})_a\text{-}20_{ret}\text{-}21\text{-}22_{call}\text{-}(_d\text{-}1\text{-}2\text{-}3\text{-}9\text{-})_d\text{-}22_{ret}\text{-})_t$
$(_t\text{-}20_{call}\text{-}(_a\text{-}12\text{-}13\text{-})_a\text{-}20_{ret}\text{-}21\text{-}22_{call}\text{-}(_d\text{-}1\text{-}2\text{-}3\text{-}5\text{-}4\text{-}3\text{-}9\text{-})_d\text{-}22_{ret}\text{-})_t$
$(_t\text{-}20_{call}\text{-}(_a\text{-}12\text{-}13\text{-})_a\text{-}20_{ret}\text{-}21\text{-}22_{call}\text{-}(_d\text{-}1\text{-}2\text{-}3\text{-}5\text{-}6\text{-}4\text{-}3\text{-}9\text{-})_d\text{-}22_{ret}\text{-})_t$
$(_t\text{-}20_{call}\text{-}(_a\text{-}12\text{-}13\text{-})_a\text{-}20_{ret}\text{-}21\text{-}22_{call}\text{-}(_d\text{-}1\text{-}2\text{-}3\text{-}5\text{-}6\text{-}4\text{-}3\text{-}9\text{-}10\text{-})_d\text{-}22_{ret}\text{-})_t$

*We use '(' to denote entry node and ')' to denote exit node.

(c) whole transaction basis paths for the two transactions

Figure 3: Illustrating example – Pool-Shark (hosted at [31])

### B. Whole Transaction Basis Path Set

Based on the program model, the execution behavior of a smart contract application is determined by a sequence of transactions. Transactions may be represented by combining the control flow graphs (CFGs) of all functions that might possibly be executed in the transactions. We call such combined graph a *transaction control flow graph* (TCFG). An example is shown in Figure 3 (b). For each function *f*, the control flow graph for *f* has a unique *entry vertex* Entry$_f$, and a unique *exit vertex* Exit$_f$. If the function is a public/external function, then its entry vertex is also marked as a *transaction entry vertex*. The other vertices present statements and predicates in the usual way, except that each function call is represented by two vertices, a *call-site vertex* and a *return-site vertex*. In addition to the ordinary intra-procedural edges that connect the vertices, TCFG also contains a *call-edge* and a *return-edge*, which connect the call-site/return-site vertices of the caller function with the entry/exit node of the callee function, respectively.

The part of TCFG introduced above is similar to the *system control flow graph* or *supergraph* [32] used in many works on inter-procedural program analysis problems. However, there are two important smart contract features that are not yet used in the previous example, and therefore demand additional constructs to handle. The example shown in Figure 4 shows these two features. This example is adapted from the real world contract attacked in the famous DAO incidence [8].

Feature 1: *Exception-handling*. EVM supports exceptions that can revert state for error handling. The function **revert** is used for flagging an error and reverting the current call. In the scenario that the exception occur in a callee function, it would be re-thrown in the caller function automatically, except when the callee function is called by low-level APIs such as **send**, **delegatecall**, **call**, and **callcode**.

```
function depositFunds() public payable {
1:     balances[msg.sender] += msg.value;
}
function withdrawFunds (uint256 withdraw) public {
2:     _withdrawFunds(withdraw);
}
function _withdrawFunds (uint256 withdraw) internal {
3:     bool result=msg.sender.call.value(withdraw)();
4:     require(result);
5:     balances[msg.sender] -= withdraw;
6:     lastWithdrawTime[msg.sender] = now;
}
```

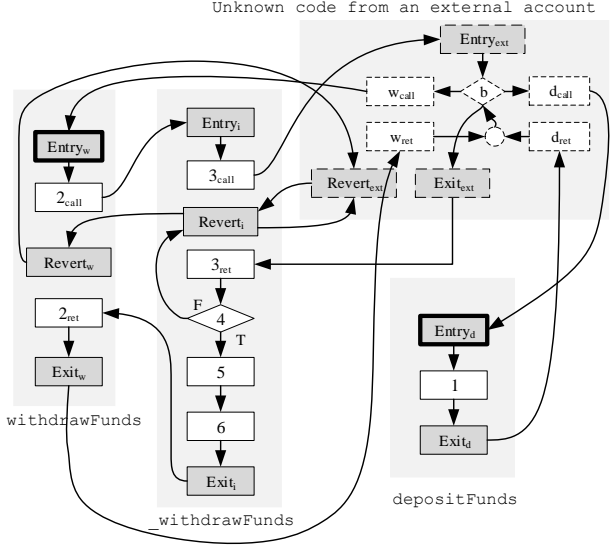

Whole transaction basis path set for `withdrawFunds`:

($_w$-2$_{call}$-($_i$-3$_{call}$-($_{ext}$-b-)$_{ext}$-3$_{ret}$-4-5-6-)$_i$-2$_{ret}$-)$_w$

($_w$-2$_{call}$-($_i$-3$_{call}$-($_{ext}$-b-)$_{ext}$-3$_{ret}$-4-Revert$_i$-Revert$_w$

($_w$-2$_{call}$-($_i$-3$_{call}$-($_{ext}$-b-d$_{call}$-($_d$-1-)$_d$-d$_{ret}$-b-)$_{ext}$-3$_{ret}$-4-5-6-)$_i$-2$_{ret}$-)$_w$

($_w$-2$_{call}$-($_i$-3$_{call}$-($_{ext}$-b-w$_{call}$-($_w$-2$_{call}$-($_i$-3$_{call}$-($_{ext}$-b-)$_{ext}$-3$_{ret}$-4-5-6-)$_i$-2$_{ret}$-)$_w$-w$_{ret}$-b-)$_{ext}$-3$_{ret}$-4-5-6-)$_i$-2$_{ret}$-)$_w$

($_w$-2$_{call}$-($_i$-3$_{call}$-($_{ext}$-b-w$_{call}$-($_w$-2$_{call}$-($_i$-3$_{call}$-($_{ext}$-b-)$_{ext}$-3$_{ret}$-4- Revert$_i$-Revert$_w$-Revert$_{ext}$- Revert$_i$- Revert$_w$

Cyclomatic Number = 28 - 25 + 2 = 5

Figure 4: A smart contract with unknown external function call and exception-handling

These low-level APIs will return false instead when an exception occurs.

In order to handle EVM exception semantics, we add a *revert node* Revert$_f$ for each function *f* that might revert itself or transitively call any function that might revert. For statements that might potentially revert (such as the **require** statement), we add a branching node to represent revert condition checking, and connect the revert branch to the revert node with a *revert edge*. In addition, we add a *cascading-revert edge* to connect the revert node in the callee function to that in the caller function.

Feature 2: *Call to unknown external function*. With the built-in type **address**, smart contract can call the function of arbitrary external contracts. For example, at line 3 in Figure 4, **msg.sender** refers to the address of the caller, which can be *any* contract or account in the Ethereum. By invoking the low-level API **call**, the fallback function of **msg.sender** is called. As the implementation of this fallback function will not be known during testing, we have to assume every possibility, including calling back into itself, other functions of the same contract, or those of yet another external contact. This features open up a lot of unexpected control flow paths that we need to incorporate into TCFG and test thoroughly.

In order to address this feature, we create a virtual sub-flowgraph to represent what might possibly happen in the unknown external function, which is illustrated at the top-right corner of the TCFG in Figure 4. Essentially, this subgraph contains a loop of calls into *any* public/external functions of the contract. As the called functions might revert, the subgraph also adds a revert node to support cascading revert.

To summarize, the definition of TCFG is as follows:

**Definition 1**: A *transaction control flow graph* (TCFG) for a smart contract *c* is a graph <*V*, *E*> where *V* is the set of vertices and *E* is the set of edges. Vertices in *V* are divided into subsets: $V_{entry}$, $V_{exit}$, $V_{expr}$, $V_{pred}$, $V_{revert}$ represent entry, exit, expression, predicate, and revert nodes; $V_{call}$, $V_{ret}$ represents call node and return node at the callsite; $V_{t\text{-}entry} \subseteq V_{entry}$ represents transaction entry node. Edges in *E* are divided into subsets $E_{flow}$, $E_{call}$, $E_{ret}$, $E_{revert}$, $E_{cascading\text{-}revert}$. $E_{flow}$ are intra-procedural edges between nodes in *V*, and $E_{call} \subseteq V_{call} \times V_{entry}$, $E_{ret} \subseteq V_{exit} \times V_{ret}$, $E_{revert} \subseteq V_{expr} \times V_{revert}$, $E_{cascading\text{-}revert} \subseteq V_{revert} \times V_{revert}$. □

**Definition 2**: A node sequence $p = (n_1, n_2, …, n_k)$ is a *whole transaction path* (WTP) in a TCFG <*V*, *E*> if *p* satisfies the following requirement: (1) $n_1 \in V_{t\text{-}entry}$ and $n_k \in V_{exit} \cup V_{revert}$; (2) $(n_i, n_{i+1}) \in E$; (3) Let $f_1, f_2, …f_w$ be the set of all functions in the contract and let *p′* be a subsequence of *p* where all nodes in *p* are removed except for those in $V_{entry}$, $V_{exit}$, and $V_{revert}$, then *p′* must match the following context free grammar:

$S ::= Entry_{f1}\ S\ Exit_{f1}\ |\ Entry_{f1}\ S\ Revert_{f1}$

……

$S ::= Entry_{fw}\ S\ Exit_{fw}\ |\ Entry_{fw}\ S\ Revert_{fw}$

$S ::= S\ S\ |\ \varepsilon$ □

The contract in Figure 3 and Figure 4 can both generate infinite numbers of WTPs. For the example in Figure 3, this is because of the loop at line 2-8. For that in Figure 4, this is due to the reentrancy at line 3 can recycle infinitely through the unknown external function. Therefore, directly defining coverage criteria for smart contracts with WTP could make testing intractable. To address this issue, we provide a solution inspired by the widely used and studied method of basis path testing proposed by Thomas McCabe [28].

The idea of basis path testing is to consider the (usually infinite) set of paths in CFG as a vector space. The *basis* of a vector space contains a limited set of vectors that are linearly independent of one another, and have a spanning property: everything within the vector space can be expressed in terms of the elements within the basis. McCabe argued that by rigorously testing the path in the (always limited) basis set, most of the defects can be exposed, therefore there is no need to test every path in the (potentially infinite) whole set.

Formally, for a control flow graph *G* with *k* edges $e_1, e_2, …, e_k$, a path *p* is represented as a vector $\langle x_1, x_2, …, x_k \rangle$, where $x_i$ is the number of time edge $e_i$ occurs in *p*. A path set *P* is

*linearly independent* if none of the path in *P* is represented by a vector that can be expressed as a linear combination of vectors of other paths. A linearly independent path set *P* is called the basis path set of *G* if every path in *G* can be expressed as a linear combination of paths in *P*.

Previously, the basis path set is only defined intra-procedurally on CFG. We now extend it to TCFG.

**Definition 3**: Given a TCFG and a transaction entry node *n*, a path set *P* is called the **whole transaction basis path set of *n***, or **WTPBS(*n*)** for short, if: 1) every path in *P* is a WTP that starts from *n*; 2) *P* is linearly independent; and 3) every WTP that starts from *n* can be expressed as a linear combination of paths in *P*.  □

Note that the set of paths that satisfies this requirement might not be unique. In this case, we can designate any of such sets as the whole transaction basis path set of *n*.

*C. Bounded Transaction Interaction*

The above definitions only involve the paths inside individual transactions. We now introduce the definition on inter-transaction control flow.

**Definition 4**: Given a *dapp* <*A, C*> and upper bound *k*, a sequence $q = \langle (a_1, c_1, f_1, o_1), (a_2, c_2, f_2, o_2), ...., (a_k, c_k, f_k, o_k) \rangle$ is called a **k-bounded transaction interaction** if: for every $(a_i, c_i, f_i, o_i)$ there exists some values of *e, I, R, L* that makes < $a_i, c_i, f_i, e, o_i, I, R, L$> be a feasible transaction of *dapp*.  □

III. COVERAGE CRITERIA

Based on the above definitions, we are ready to define the coverage criteria for smart contract testing. Firstly we shall define the basic coverage requirement:

**Definition 5**: Given a *dapp* and upper bound *k*, a **k-bounded transaction coverage requirement** is defined as a tuple (*q, w*), where $q = \langle (a_1, c_1, f_1, o_1), ...., (a_k, c_k, f_k, o_k) \rangle$ is a k-bounded transaction interaction, and *w* is a set of paths $p_1,..., p_k$ such that for *i*=1, 2, ... *k*, $p_i \in$ WTPBS(Entry$_{f_i}$) and ends with Exit$_{f_i}$ if $o_k$ is *success*, or Revert$_{f_i}$ if $o_k$ is *revert*.  □

A test case of a smart contract application *dapp* <*A, C*> is defined as a tuple ($\phi, \delta$), where $\phi$ maps each external account *a* in *A* to its balance in Ethers and $\delta$ is an input sequence $\langle (a_1, c_1, f_1, e_1, I_1), ..., (a_n, c_n, f_n, e_n, I_n) \rangle$. Each item ($a_i, c_i, f_i, e_i, I_i$) defines the input for a transaction: $a_i$ is the external account, $c_i$ is the contract, $f_i$ is the external/public function, $e_i$ is Ethers sent with the transaction, and $I_i$ is the message call data. All of them can be either constant or calculated from transaction outputs, logs, view/public function call return values from previous transactions execution results. A test case is valid if the input sequence can be executed on Ethereum from a state where none of the contract in *C* have been deployed (that is, a clear state).

Intuitively, given a *dapp* <*A, C*>, we said that a test case *s* covers a k-bounded transaction coverage requirement *r*= (*q, w*) if we execute *s* on a clear state, and the Ethereum blockchain generates a sequence of transactions $\langle T_1, T_2, ... T_n \rangle$ from which we can find a substring $\langle T_i, T_{i+1}, ... T_{i+k-1} \rangle$ that matches the transactions and basis paths specified in *r*. A

---

**Input**: a TCFG <*V, E*>, a transaction entry node *n* in $V_{t\text{-entry}}$
**Output**: WTPBS(*n*)
1: Find *p* as the WTP that starts with *n* with the least predicate nodes
2: *P* = {*p*}
3: $V' = \{v \in V \mid v \text{ is reachable from } n\}$, $E' = \{(v_1, v_2) \in E \mid v_1, v_2 \in V'\}$
4: while $|P| < |E'| - |V'| + 2$  //Cyclomatic Complexity number
5:     Let $p = (n_1, n_2, ..., n_k)$, find the smallest *i* such that ($n_i, n_{i+1}$)
6:     has not occurred in any path in *P*.
7:     Find another WTP $p'$ with the prefix ($n_1, ..., n_i, w$), $w \neq n_{i+1}$
8:     if no such $p'$ is found, then break
9:     $P = P \cup \{p'\}$, $p = p'$
10: return *P*

Algorithm 1: Whole transaction basis path set generation

**Input**: a *dapp* <*A, C*> and the bound *k*
**Output**: all k-bounded transaction coverage requirements
1: $R = \varnothing$
2: *U* = the set of all valid 4-tuples (*a, c, f, o*) where $a \in A, c \in C$,
3:     $f \in$ public/external functions of *c*, $o \in \{success, revert\}$
4: *Q* = all length-*k* permutations on *U*
5: while $|Q| \neq 0$
6:     pick a permuntation *q* and remove it from *Q*
7:     for each ($a_i, c_i, f_i, o_i$) in *q*, pick a path $p_i$ from WTPBS(Entry$_{f_i}$)
8:     $p_i$ must end with Exit$_{f_i}$ if $o_i$ is *success*, or Revert$_{f_i}$ if $o_i$ is *revert*
9:     enumerate all such $\langle p_1, p_2, ..., p_k \rangle$, add it into *R*.
10: return *R*

Algorithm 2: k-bounded transaction coverage requirement generation

coverage requirement is feasible if there exists at least one test case that can cover it.

**Definition 6**: Given a *dapp* and upper bound *k*, a set of smart contract test cases *S* achieves **k-bounded transaction coverage criteria** if for every feasible k-bounded transaction coverage requirement *r*, there exists at least one test case *s* in *S* such that *s* covers *r*.

With the different values of *k*, we can obtain a family of coverage criteria with increasing numbers of coverage requirements, and therefore require increasing numbers of test cases to satisfy. When the value of *k* reduced to 1, it becomes essentially whole transaction basis path coverage without considering the interaction between transactions.

IV. ALGORITHMS

Testing a smart contract application with k-bounded transaction coverage criteria essentially involve four steps: 1) enumerate all coverage requirements; 2) generate at least one test case to cover each feasible coverage requirement *r*; 3) implement a test oracle to check test outcome correctness; 4) run every test case to find potential failure. In this section we introduce the algorithms for step 1, while the algorithms on test data generation and test oracle are put in future work.

Algorithm 1 shows how to generate whole transaction basis path set. The idea is to start with a baseline WTP, then vary exactly one decision outcome to generate each successive WTP until the size of the path set reach Cyclomatic Complexity number. The proof on the correctness can be derived from the proof in [28]. We omit it due to a lack of space.

Algorithm 2 shows how to generate all $k$-bounded transaction coverage requirements. Essentially, the algorithm is to enumerate all possible length-$k$ permutation of transactions, and the enumerate all possible paths for each transaction from WTPBS($n$) we derive from Algorithm 1.

## V. CASE STUDY

We perform a case study on a real world smart contract application Pool-Shark hosted at [31]. The whole application consists of 12 contracts with 19 functions in total, among which 9 are public/external functions that can be a transaction entry. We choose Pool-Shark because its scale and complexity is representative of common applications hosted on Ethereum [33], [34]. To facilitate the repeat of our case study, we publish all the faulty versions and test cases at [35].

In the case study, we are mainly interested in two research questions:

**Q1**: Are the $k$-bounded transaction coverage criteria more *effective* than conventional code coverage criteria, in the sense that test suites satisfying the former can detect more bugs than those satisfying the latter?

**Q2**: Are the $k$-bounded transaction coverage criteria more *efficient* than random testing as a baseline, in the sense that a test suite satisfying our coverage criteria can detect more bugs than a test suite of the same size, but generated randomly?

In order to address these two questions, we used the sufficient mutation operators [36] (such as operator-replacement, variable-replacement, and statement-omission) to seed 22 mutation faults into the source code of Pool-Shark, producing 22 faulty versions with one fault in each version. As the test requirements quickly explode with an increasing bound $k$, we set the value of $k$ as 2.

As mentioned in Section IV, the whole testing process consist of four steps. In step 1, we apply the algorithms defined in Section IV to enumerate all the coverage requirements. With the value of $k$ as 2, Pool-Shark generates 729 coverage requirements in total. In step 2, we first use symbolic execution to solve the path constraints for each coverage requirement. The objective is to derive the contract state that can trigger a transaction sequence that cover the requirement. Next, we manually construct a transaction sequence that can get to this state. By combining the two sequences we get a complete test case. In step 3, we record the transaction execution outcome, event logs, and the return values of all pure/view functions of the correct version as test oracle. In step 4, the Truffle framework [37] is used to execute the test script. This process is repeated for statement coverage testing. The only difference is in step 2, where the coverage requirement is changed to every executable statement in Pool-Shark.

With the above process, we generate 81 test cases for $k$-bounded transaction coverage criteria and 9 test cases for statement coverage criteria. Note that the number of test cases are significantly less than the number coverage requirements. It is due to the fact that one test case can cover multiple coverage requirements. Moreover, not all of the coverage requirements are feasible. To compare with random testing, we also construct a random test suite by randomly generating a transaction sequence with random message call data.

Table 1: case study result

| Testing method | #. of logic faults detected | %. Of detected faults |
|---|---|---|
| $k$-bounded transaction coverage | 20 | 90.9% |
| Statement coverage | 8 | 36.3% |
| Random | 16 | 72.7% |

The comparison result is shown in Table 1. It can be observed that the $k$-bounded transaction coverage testing is significantly more effective than statement coverage testing, as it detects nearly 55% more faults. In fact, all but two of the faults are detected by our testing technique. After manual examination, we find out that these two faults are indeed equivalent mutants and therefore, there is no test case that can detect them. At the same time, the $k$-bounded transaction coverage testing is more efficient than random testing: with the same number of test cases, it detects 18.2% more faults. This suggests our proposed criteria can be of practical value to smart contract developers.

## VI. RELATED WORK

Existing vulnerability detection approaches can be classified according to their underlying techniques. Some of them rely on static program analysis. For example, systems including Oyente [38], Maian [11], Teether [12], Gasper [13] and the work by Grossman et al. [39] use symbolic execution to explore whether there exists paths that can trigger any known vulnerability, while ContractFuzzer [40] uses random fuzzing to find vulnerability instead. Other works rely on formal verification tools. For example, Zeus [14] uses abstract interpretation and constrained horn clauses, Vandal [41] uses Datalog theorem prover, and Grishchenko et al. [42] use F* theorem prover. Our work is complementary to these works.

There is a plenty of work in the literature on testing coverage criteria [43]. Those that are widely referred to include the control flow coverage criteria [21], dataflow coverage criteria [22], logic coverage criteria [44], interface coverage criteria [24], and mutation score [36]. Our proposed $k$-bounded transaction coverage criteria share some of the ideas in the basis path coverage and interface coverage criteria. The merit of our work is that we extend these ideas to address the unique characteristics of the smart contract program model.

## VII. CONCLUSION

The importance of smart contract testing has been recognized, but there is a lack of research on how to systematically test smart contract application. In this paper, we analyze the unique characteristics of Ethereum smart contract model and propose the notions of *whole transaction basis path set* and *bounded transaction interactions*. Based on these two notions, we define $k$-bounded transaction coverage criteria for smart contract testing. We conduct an experiment to study its effectiveness. The initial results show that testing based on *k*-bounded transaction coverage criteria can be more effective than the conventional testing methods such as statement coverage testing and random testing. In future work, we will address the test generation problem that has not been covered in this paper.